# Mechanisms and Scale-up Potential of 3D Solar Interfacial-Evaporators

James H. Zhang, Rohith Mittapally, Abimbola Oluwade, Gang Chen[*]

Department of Mechanical Engineering

Massachusetts Institute of Technology

Cambridge, MA 02139

**Abstract**

Evaporation rates from porous evaporators under sunlight have been reported to exceed the solar-thermal limit, determined by relating the incoming solar energy to the latent and sensible heat of water, for applications in desalination and brine pond drying. Although flat two-dimensional (2D) evaporators exceeding the solar limit implies a non-thermal process, tall three-dimensional (3D) solar evaporators can exceed it by absorbing additional environmental heat into its cold sidewalls. Through modeling, we explain the physics and identify the critical heights in which a fin transitions from 2D to 3D evaporation and exceeds the solar-thermal limit. Our analyses illustrate that environmental heat absorption in 3D evaporators is determined by the ambient relative humidity and the airflow velocity. The model is then coarse-grained into a large-scale fin array device on the meters scale to analyze their scalability. We identify that these devices are unlikely to scale favorably in closed environment settings such as solar stills. Our modeling clearly illustrates the benefits and limitations of 3D evaporating arrays and pinpoints design choices in previous works that hinder the device's overall performance. This work illustrates the importance in distinguishing 2D from 3D evaporation for mechanisms underlying interfacial evaporation exceeding the solar-thermal limit.

**INTRODUCTION**

Passive solar evaporation to separate water from dissolved minerals has the potential to be a low capital cost and green method to produce clean water and harvest critical minerals. Interfacial solar evaporating materials with capillary wicking abilities has been shown to be an especially promising strategy due to its ability to concentrate the solar energy in a thin interfacial region near the evaporating surface.[1–7] Under standard one sun insolation of 1000 W/m$^2$ and assuming all of solar energy is used for evaporation, one arrives at a maximum evaporation rate of around 1.49 kg/m$^2$-hr, which we call the solar-thermal limit.

However, many studies have reported evaporation rates that exceed this solar-thermal limit.[6,8–17] For 2D solar-driven interfacial evaporating materials in which the solar absorbing area is nominally the same as its evaporating area, reports have demonstrated evaporation rates beyond the solar-thermal limit by 2 to 4 times.[6,8–12] Our recent work shows that such high evaporation rates imply water evaporates in the form of clusters, i.e. super solar-thermal, because no region of the evaporation system is below the ambient temperature and the reduced latent heat hypothesis is incorrect.[18,19] Although the details of this phenomenon are still under investigation, our group has interpreted such super solar-thermal evaporation as arising from photons directly cleaving off water molecular clusters, which we called the photomolecular effect.[8,18,20–22]

---

[*] Email: gchen2@mit.edu

For tall evaporators, which we call 3D solar interfacial-evaporators and will quantitatively define them later, the evaporating surface area is much larger than its solar absorbing area due to its extended surface similar to fins used in heat transfer devices.[13,15,23–27] It has been well appreciated in the field that these structures can exceed the solar-thermal limit due to the structure absorbing additional environmental energy along its sidewalls not exposed to direct sunlight.[14] The reported evaporation rates normally range between 3 to 5 times the solar-thermal limit[26,27] and sometimes even up to 7 times using forced convection,[17] based on the projected top cross-sectional area of the absorber. Such high evaporation rates have attracted lots of materials development and lab prototype testing. However, very few reports have analyzed the scale-up potential of 3D evaporators beyond a few fins on the decimeter device scale nor mechanistically studied the physics of the device.[14,17,28–30] Experiments from Chen et al.[16] showed as the array of extended surfaces increases, the evaporation rate per structure decreases, illustrating the challenges of scaling up 3D solar evaporators. Yang et al.[31] hypothesized that the vapor from the solar absorbing region might re-condense in the evaporative cooling region due to the vapor concentration difference, leading to degraded performance with taller fin heights inside the condenser chamber. Although many works have analyzed 3D evaporators through heat transfer equations,[13,14,25,27] very few have explicitly considered vapor transport which drives the environmental heat absorption.[17,32] Currently, there lacks a systematic study to help guide these discussions for 3D solar-interfacial evaporators that considers both heat and mass transfer kinetics. No criteria have been established to clearly distinguish 2D from 3D effects, which is important for studying mechanisms behind solar-interfacial evaporation exceeding the solar-thermal limit.

In this work, we will systematically analyze the performance of 3D solar interfacial-evaporators, starting from a single fin and extend to scaled-up systems for both forced and natural convective conditions. First, a simplified model will be constructed to illustrate the performance of a single 3D solar evaporator and reveal the underlying physics. Criteria will be established to demarcate when the evaporating structures can be treated as 3D and when it will reach or exceed the solar-thermal limit. Then, the model will be coarse-grained to a large solar device to study its scale-up potential in forced convective conditions. Our model illustrates that environmental heat input can only occur if the ambient air is below 100% relative humidity (RH) and the rate of environmental heat input depends on the airflow velocity and the RH. Our analysis shows that despite the high performance of a single and a few fins, the performance of large-scale 3D solar evaporator arrays will degrade significantly in low airflow regimes and in closed environments such as in solar still devices because of limitations in vapor transport kinetics.

## RESULTS AND DISCUSSIONS

### Mechanisms and performance of a single fin

The performance of a 3D solar evaporator is driven by the balance between evaporative cooling effects from vapor transport and heat transport kinetics from the ambient environment. We will consider a cylindrical shaped 3D solar evaporator, often called a pin shaped fin, evaporating into a large ambient reservoir. In a typical laboratory setting, the fin is inserted into a larger diameter container of water (Fig. 1a). For now, we assume the excess surface area of the container is covered and the solar beam spot size exactly matches the fin's cross-sectional area. This causes the top surface to be elevated in temperature $T$ and have a high saturated vapor mole fraction $c_{v,s}$. The sidewalls that are not exposed to direct sunlight drops below the ambient temperature due to evaporative cooling effects, leading to heat absorption from the environment to sustain further evaporation. The solar-thermal limit for 2D evaporation can be calculated by assuming all solar energy ($q''_{sun}$) is used to heat up and evaporate water, i.e.,

$$\dot{m}''_{sun,max} = \frac{q''_{sun}}{h_{fg} + c_p \Delta T_s} \quad (1)$$

where $h_{fg}$ is the latent heat of water, $c_p$ is the specific heat of water, and $\Delta T_s$ is the temperature rise of the surface. The above expression gives values between 1.45 to 1.49 kg/m²-h depending on the evaporating surface temperature of the solar absorbing region.

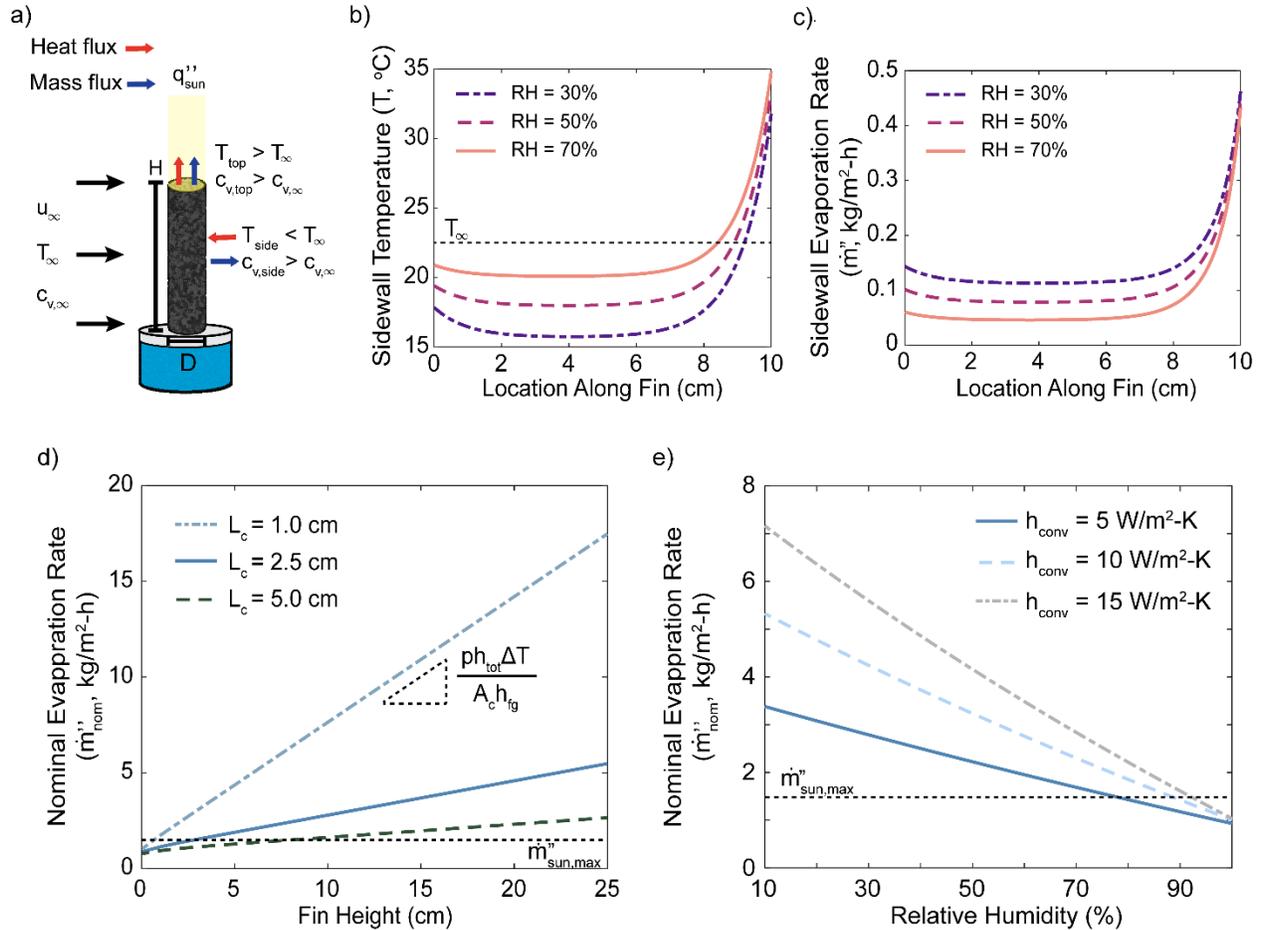

**Fig. 1** Performance and mechanism of single 3D evaporating fin. a) Diagram of heat transfer, mass transfer, and typical testing geometry in laboratory test. Unless stated otherwise, the studied base case is a fin with height ($H$) of 10 cm, a diameter ($D$) of 2.5 cm, and a thermal conductivity ($k_m$) of 0.3 W/m-K evaporating into an ambient at ($T_\infty$) 23 °C, 30% relative humidity (RH), and a convective sidewall heat transfer coefficient ($h_{conv}$) of 5 W/m²-K. The RH and ambient temperature determine the ambient vapor molar concentration $c_{v,\infty}$. Predicted sidewall b) temperature profile and c) local evaporation rates for different RH. d) Predicted nominal evaporation rate's dependence on fin geometry. e) Predicted nominal evaporation rate as a function of RH and $h_{conv}$.

**Single fin model and metrics.** We constructed an ideal model for a single fin, including solar heating on top surface, heat conduction along the fin, and evaporative, radiative, and convective heat exchange with the ambient (see Methods section). Unless otherwise noted, the performances are calculated using a base case scenario in which the extended surface has a diameter ($D$) of 2.5 cm, a height ($H$) of 10 cm, a thermal conductivity ($k_m$) of 0.3 W/m-K, evaporating into an ambient at ($T_\infty$) 23 °C, 30% relative humidity (RH), and an external sidewall convective heat transfer coefficient $h_{conv}$ of 5 W/m²-K. We will analyze the results in a forced convection setting with bulk air velocity $u_\infty$ in crossflow to the cylinder to decouple the heat transfer coefficients from the surface temperatures but expect similar results for natural convection. For a given convective condition, the heat transfer coefficient $h_{conv}$ is interrelated with the vapor mass transfer coefficient $g_{conv}$ because of similarities in the boundary layers above the surface.[33]

We will define the nominal evaporation rate, $\dot{m}''_{nom}$, as the total evaporation rate of the fin normalized to only the top projected cross-sectional area $A_c$. The nominal evaporation rate is the metric commonly reported in previous literature to characterize their performance and governs the solar absorption area when light is incident only on the top surface.

$$\dot{m}''_{nom} = \frac{\int \dot{m}'' dA_{wetted-surface}}{A_c} \quad (2)$$

Later, we will further discuss evaporation rate normalized to the device footprint area.

**Performance of a single fin.** Fig. 1b illustrates the predicted surface temperature along the fin's sidewalls for different RH values, illustrating that the constructed model captures the physics of 3D evaporators shown in Fig. 1a. The top evaporating surface is above the ambient temperature due to solar absorption. About 1.5 cm below the hot top surface, the extended surface drops below the ambient temperature from evaporative cooling. In the middle region along the sidewall, the temperature profile is flat, agreeing with previous experiments.[13,14,27] The flat temperature value is strongly determined by the ambient RH. Near the bottom of the extended surface, the temperature increases again due to the heat transfer with the water reservoir beneath the fin. Fig. 1c shows the local evaporation flux along the fin's sidewall for the same given conditions. The local sidewall evaporation rate from the flat temperature region is between 73% to 89% smaller than the hot solar absorbing region near the top because the saturated vapor concentration decreases rapidly with the surface temperature. The solar absorbing "hot" top region has much faster evaporation kinetics than the evaporatively cooled "cold" regions along the middle region.

Using the model, we can predict the nominal evaporation rate's dependence on the fin height and diameter (Fig. 1d). The model predicts that the nominal evaporation rate increases linearly with the fin height due to the increased evaporating surface area assuming no dry out occurs. In a real system, there will be a limit to the height of the fin based on the capillary water pumping in the designed fin. The reason why the nominal evaporation rate increases linearly with fin height is explained with the temperature profiles found previously. As the fin gets taller, it has minimal effects on the temperature profiles near the hot solar-absorbing top and the cooler bottom exchanging heat with the water reservoir below. These features are determined by the balance of the fin's heat conduction along its length with the heat and vapor exchange with the ambient reservoir at the respective ends. Increasing the fin's height only increases the length of the flat temperature profile region in the middle. The energy balance in the flat temperature region is a simple relationship between the evaporative cooling from vapor transport kinetics and the combined convective and radiation heat transfer coefficient with the ambient.

$$g_{evap} C_g h_{fg} \left( c_{v,s}(T) - RH c_{v,s}(T_\infty) \right) = h_{tot}(T_\infty - T) \tag{3}$$

where $C_g$ is the molar density of air, $h_{fg}$ is the latent heat of vaporization per mole of water, and $h_{tot}$ is the combined convective and radiative heat transfer coefficients. Eq. (3) illustrates that the properties of the flat temperature region is independent of the fin's properties beyond its blackbody emissivity and driven by the air side boundary layers.[33] At high airflow external velocities when radiation becomes negligible relative to convection, this temperature should approach the wet-bulb temperature.

The performance of 3D evaporators and its environmental heat absorption is governed by the ambient RH near the fin and the external airflow velocity. If the ambient is near 100% RH, Eq. (3) will simplify to the null solution and no environmental heat absorption can occur. Using this knowledge, we can estimate that the slope governing the increase in the nominal evaporation rate with the fin height is

$$\frac{d\dot{m}''_{nom}}{dH} = \frac{p h_{tot} \Delta T}{A_c h_{fg}} \tag{4}$$

where $\Delta T$ is the temperature difference between the flat region and the ambient from Eq. (3) and $p$ is the perimeter of the fin. This linear growth in the nominal evaporation rate with height agrees well with a previous experimental study.[30]

Due to the strong dependence of the nominal evaporation rate on the ambient RH and convective conditions, we further plot the nominal evaporation rates as a function of these two variables (Fig. 1e). As the sidewall convective heat and mass transfer coefficient increases from 5 to 15 W/m$^2$-K, the nominal evaporation rates increase by over 2 times at 10% RH and are above the solar-thermal limit due to environmental heat absorption. However, as the ambient humidity increases to 100% RH, all curves predict that the performance will degrade to below the solar-thermal limit for 2D structures and converge towards similar performances. At 100% RH, the 3D evaporator will always behave like a 2D evaporator regardless of its height.

**Critical heights for 3D evaporators**

There is a critical height in which the evaporator will transition from a 2D to a 3D evaporator that can absorb environmental heat. If the water reservoir beneath the fin is the same temperature as the ambient, we can identify this critical height, $H_{cr,2D}$, as the point in which the fin's temperature profile will drop below the ambient temperature due to evaporative cooling and start to absorb environmental heat (Fig. 2a). When the fin behaves like a 2D evaporator, the temperature profile will monotonically decrease from the hot evaporating top surface to the ambient bottom temperature. This suggests that the $H_{cr,2D}$ will be related to the fin parameter $\beta$.

$$\beta = \sqrt{\frac{h_{conv} p}{k_m A_c}} \quad (5)$$

When the convective heat transfer coefficient is at 5 W/m²-K and the ambient is 50% RH, the non-dimensional critical height is about 0.85. Using the base case scenario for a diameter of 2.5 cm and ambient sidewall heat transfer coefficient of 5 W/m²-K, this corresponds to a height of about 1.15 cm. For structures with much smaller cross-section areas, this critical height is significantly lower.[26] Fins with heights below $H_{2D}$ can only lose heat to the environment whereas fins taller than $H_{2D}$ can absorb heat from the environment. As the RH approaches 100%, $H_{2D}$ diverges and can never absorb environmental heat. Higher convective heat transfer coefficients decrease $H_{2D}$ because the evaporative and convective cooling will remove the solar heat more rapidly. The critical height is not just a function of the nondimensional parameters due to the nonlinearity of the saturated vapor molar concentration dependence on temperature.

From the discussion on 2D to 3D evaporator transition, it will be meaningful to further identify how tall a fin needs to be, $H_{cr,th}$, for the nominal evaporation to reach the solar-thermal limit (Fig. 2b). $H_{cr,2D}$ doesn't determine the critical height needed to exceed $\dot{m}''_{sun,max}$ because the fin first needs to first absorb sufficient environment heat to compensate the heat lost from convective cooling at the top solar absorbing region. Rather than normalizing to the fin parameter $\beta$, $H_{cr,th}$ is normalized to the fin's geometric aspect ratio $p/A_c$ because the flat-temperature middle region that absorbs solar energy is independent of the fin's properties if dry-out doesn't occur. As the ambient RH increases, $H_{cr,th}$ increases exponentially because the environmental heat absorption and evaporation are determined by the ambient RH. For the same reason with $H_{cr,2D}$, $H_{cr,th}$ will also diverge as the ambient humidity approaches 100%. Many previous experiments have fins with fin aspect ratios $Hp/A_c$ on the order of 10 and testing in environmental RH between 30-50%. Fig. 2b suggests that these will achieve super solar-thermal evaporation rates.

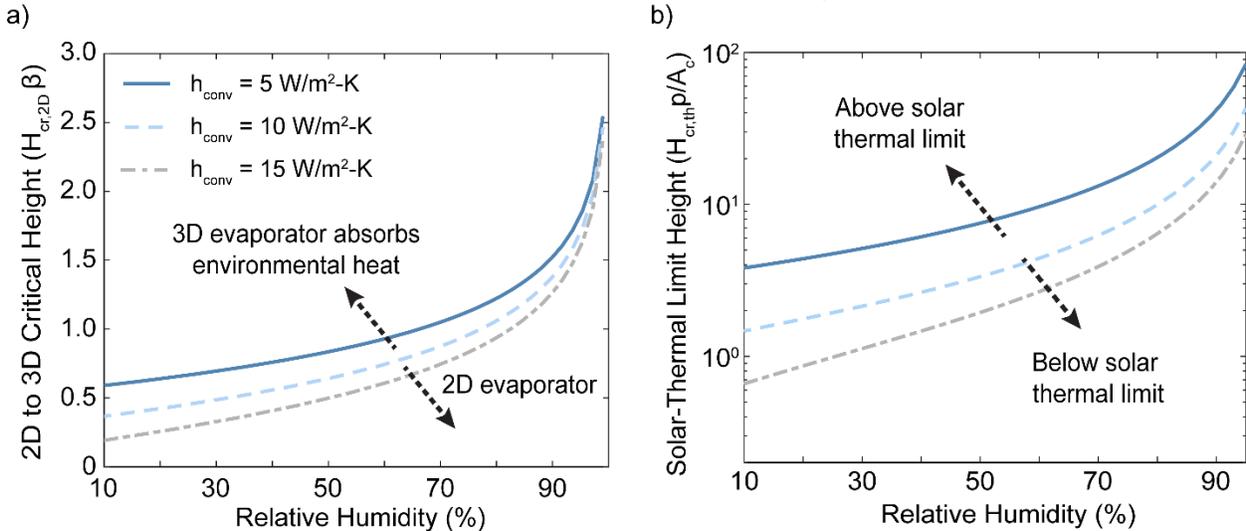

**Fig. 2** Critical heights of 3D evaporators. a) Height $H_{cr,2D}$ is when a 2D evaporator will behave like a 3D evaporator and absorb environmental energy. b) Height $H_{cr,th}$ of a 3D evaporator is when it will absorb enough environmental heat to exceed the solar thermal limit. Each curve corresponds to a different convective condition. The plots were produced use a fin with a diameter of 2.5 cm and a thermal conductivity of 0.3 W/m-K evaporating into an ambient at 23 °C.

**Performance of scaled-up array in outdoor setting**

In a scaled-up array of 3D fins for outdoors experiments, many conditions are significantly different from the laboratory setting (Fig. 3). The total solar energy and solar absorbing area of the device varies with the time of day due to the changing solar zenith angle $\theta$. The fins will absorb additional sunlight along its sidewalls, cast a shadow, and shade the projected fin's area behind it.[25] For small devices where fins have much taller heights than the device width, the device will absorb significantly more solar energy than the device area footprint, $A_{dev}$. This is because the shaded area will extend beyond the device's footprint. The device's footprint area can be related to device width and length as

$$A_{dev} = (S_l N_{rows})(S_t N_{cols}) \qquad (6)$$

where $N_{rows}$ is the number of rows of fins in the y-axis, $N_{cols}$ is the number of columns in the transverse x-axis, $S_l$ is the spacing length between rows, and $S_t$ is the spacing width between columns. The total length of the device $L$ is simply $S_l N_{rows}$. This additional area that absorbs sunlight will become negligible as the device is scaled-up in size where $L \gg H$ because this is only an edge effect of the backmost row of fins, leading to the total solar energy incident the device to be approximately equal to

$$q_{sun} = q''_{sun} A_{dev} \tag{7}$$

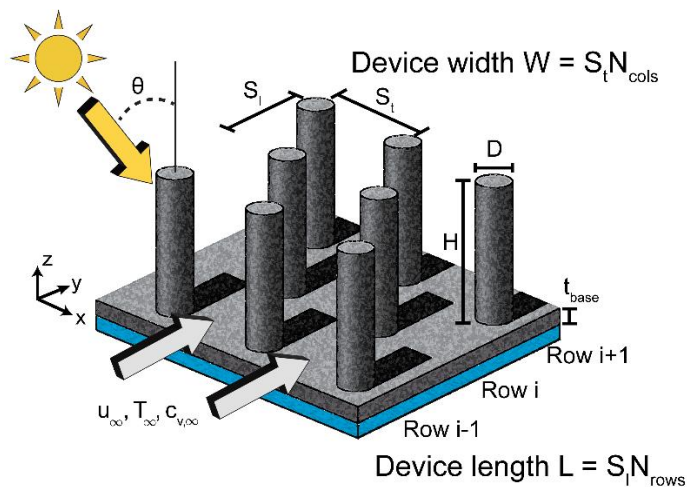

**Fig. 3** Diagram of scaled-up fin array device in outdoor setting. The time of day will change the solar absorbing area of the device due to the zenith angle $\theta$. The total device length L and width W depends on the number of fin columns and rows in the array. The geometric dimensions of the fins and air movement in crossflow to the fins are illustrated as well.

In many previously reported outdoor devices, the base of the device is a non-evaporating and non-absorbing insulating material such as white foam instead of an interfacial evaporator. We will show later that this configuration will degrade the true performance of the overall device because a significant amount of solar energy will be wasted when it can be used to further evaporate water. Finally, we define a device evaporation rate that is normalized to the device footprint area to give a representative metric for scaled-up devices that considers the solar absorption of the entire structure.

$$\dot{m}''_{dev} = \frac{\int \dot{m}'' dA_{wetted-surface}}{A_{device}} \tag{8}$$

This metric is only accurate either in the condition when $L \gg H$ or when $\theta = 0$, the latter represents best scanerio, and hence our discussion leads to upper limit in evaporation rates.

**Forced convection device model.** The transport kinetics of vapor through the device determines the performance of the fin array because the environmental heat absorption depends on the air remaining below 100% RH. As the air flows through the fin array device, it will progressively become more humid and diminish the performance of the fins further downstream. If the device becomes too large, the hotter humid air can cause vapor recondensation on the sidewalls and decrease its overall evaporation rate. These effects suggest that given a total device length, solar intensity, external airspeed, and reservoir conditions there will be an optimum fin sizing and spacing. However, the large mismatch in length scales between the boundary layer thickness and the device size make FEA simulations of an entire fin array very computationally expensive.

Using the single fin model, we developed a coarse-grained model based on a control volume analysis of the airflow (see Methods for details). In this model, we draw a series of control volume along the y-axis of the array. Each control volume along the y-axis has dimensions $S_t$ and $S_l$ such that only one fin is inside. We assume that air only flows along the y-axis such that each fin in the same row behaves identically. We will consider a fin with the same characteristics as the single fin analysis. The transverse spacing $S_t$ is 4D (10 cm) and the longitudinal spacing $S_l$ is 2D (5 cm). This will cause the extended surface to have about 85% more evaporating surface area than the base plate around it in each control volume. The incoming air is initially at 23 °C and 30% RH. For solar absorption calculations, we will assume that the zenith angle is 0° and that the incoming solar intensity is 1000 W/m², all absorbed where the solar ray lands (on the top of the fin and the base plate around the fin). The total area, $A_{base}$, that absorbs sunlight in each control volume would simply be the product of the two array spacings. This should correspond to the maximum evaporate rate since in this case, the side wall is not heated directly by the solar radiation.

In addition to the total device evaporation rate $\dot{m}''_{dev}$, we will further define a local device evaporation rate to illustrate the performance of each row of fins, denoted by index i.

$$\dot{m}''_{dev,i} = \frac{\int \dot{m}''_i dA_{wetted-surface}}{A_{base}} \tag{9}$$

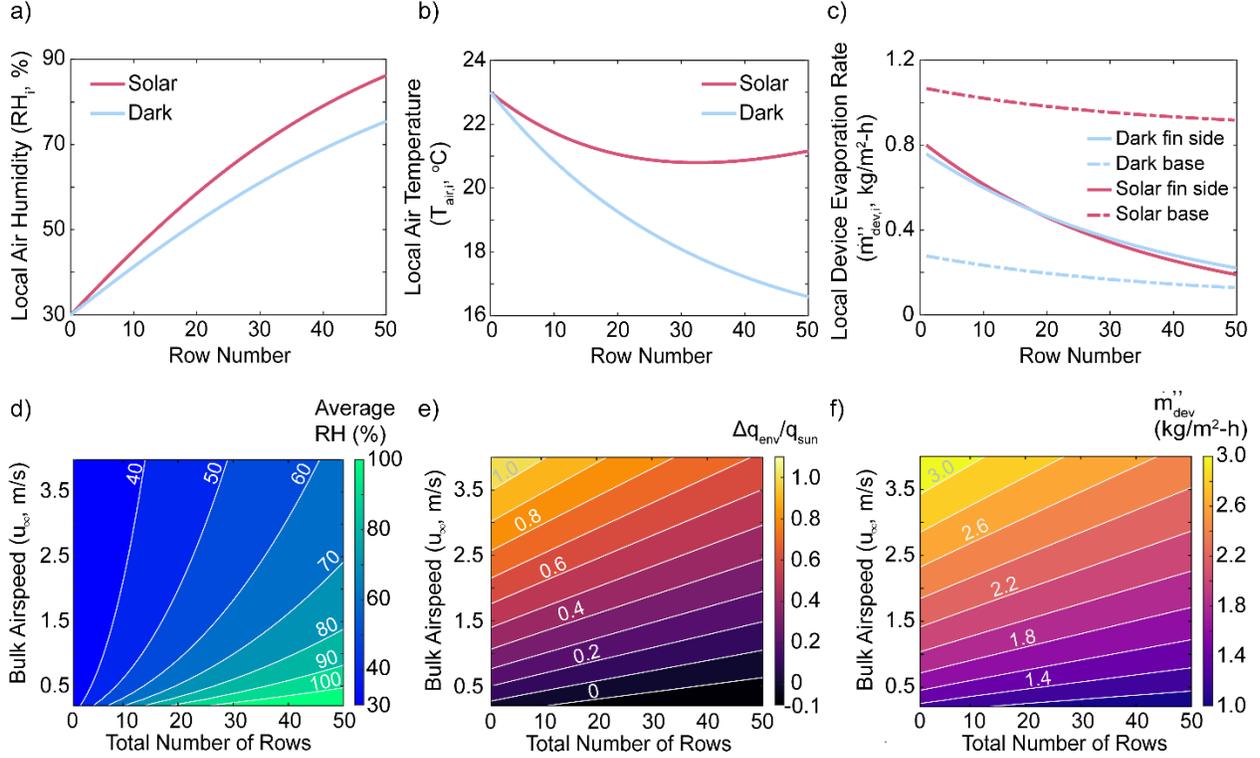

**Fig. 4** Heat and vapor exchange with airflow through the device for large scale systems. a)-c) Properties and performance of a device 50 rows (2.5 m) long with an external bulk airspeed at 1 m/s. Local air a) RH and b) temperature for dark and solar conditions as a function of row number. c) Contributions from the base plate and the fins' sidewalls to the local device evaporation rate. d)-f) Performance of devices with different lengths and subjected to different bulk airspeeds. d) Average RH of air throughout entire device. e) Ratio of total environmental heat absorption to total solar absorption. f) Total device evaporation rates. The fins have a diameter of 2.5 cm, height of 10 cm, thermal conductivity of 0.3 W/m-K, transverse fin spacing of 10 cm, and longitudinal fin spacing of 5 cm. The incoming air is initially at 23 °C and 30% RH.

where $A_{base}$ is equal to $S_t S_l$. These parameters will depend on what the local air temperature $T_i$ and humidity $RH_i$ at the specified row of fins in the device.

**Performance of device.** We can now analyze the detailed airflow properties to better understand the mechanisms of the device. Fig. 3a illustrates the local air humidity along the length (y-axis) of the device with 50 rows in width (2.5 m) for both solar and dark conditions. The air's RH initially increases rapidly and then its growth decays further downstream in the device. The RH doesn't increase as rapidly in dark conditions as in solar conditions. Environmental heat exchange between each row and the airflow will cause the local air temperature to change as well (Fig. 3b). For the solar condition, the air temperature initially drops, reaches a local minimum, and then increases afterwards. For dark conditions, the local air temperature monotonically decreases.

The reason for these trends is revealed in Fig. 3c, illustrating the contribution of the solar absorbing base plate area and the fin's sidewall contributions to the local device evaporation rate per row. For fins in both dark and solar conditions, the rows near the leading edge of the device contribute a significant amount of evaporation due to the low humidity in the air. In fact, their performances are almost identical because their evaporation rates are primarily determined by 3D environmental heat absorption effects. This causes the local air temperature to cool down due to the fins' sidewall environmental heat absorption. For the base plate near the leading edge, it will heat the air due to convective heat loss in solar conditions and cool the air in dark conditions due to an evaporative cooling effect like the fins' sidewalls. Further downstream inside the device, the RH will increase and the fins' sidewall evaporation becomes less effective, leading to the local fin sidewall evaporation rates to degrade rapidly. The base plate downstream in solar conditions continues to both evaporate into and heat up the air due to its enhanced temperatures. In dark conditions, the base area's local device evaporation flux also decreases rapidly downstream. The balance between the fins' sidewall heat gain and the base area's heat loss coupled to the changing RH and local evaporation rates create the air temperature minimum under solar conditions. In the initial rows of the device, the local device evaporation rate exceeds the solar-thermal limit due to the low RH. Further downstream as the air saturates, the local device evaporation rate degrades to below the solar-thermal limit.

These calculations illustrate that the common procedure in previous works, in which the base area is covered with a non-evaporating and non-absorbing material such as white foam, will degrade the device performance significantly. The fins will only have a high device evaporation rate if the local air RH is below 100%. In contrast, the base plate will continue to evaporate water because it heats up from solar absorption, leading to higher saturated vapor molar concentration and enhanced evaporation kinetics. Not utilizing the total solar energy on the entire device will lead to large reductions in performances and this behavior is evident if the device performance is normalized to the entire area footprint rather than only the fin's cross-sectional area.

Our single fin analysis shows that environmental heat gain can only occur if the air flow inside of the device is below 100% RH. Fig. 3d illustrates the average RH of the air across the entire device as a function of its total number of rows (length) and external airspeeds in solar conditions. At low airspeeds, the air will rapidly saturate as the device gets wider. At a bulk airspeed of 2.5 m/s, the average RH is 70% if the device has 50 rows (2.5 m) in length. This scaling relation is because the rate at which air is replaced in the device scales linearly with the bulk airspeed and the rate of evaporation scales sub-linearly with the bulk airspeed due to the Sherwood number relationship (see Methods section below). From this analysis, we can appreciate that larger devices and greater environmental gain can be achieved only if the external bulk airspeed is high and the ambient air is dry. Fig. 3e shows the total environmental heat gain $\Delta q_{env}$ relative to the total solar energy $q_{sun}$ the device absorbs. At lower airspeeds and longer device lengths, the device will lose heat to the environment ($\Delta q_{env} < 0$) because the hot base loses heat faster than the sidewalls can absorb heat. This is because the humid air cannot be replaced by dry air fast enough to enhance the fins' sidewall evaporation. The corresponding device evaporation rates are illustrated in Fig. 3f, which can exceed the solar-thermal limit under certain conditions. For a given bulk airspeed, there is an optimum number of fins and length the device can be scaled to. Due to the diminished performance of the fins as the device gets larger, the device performance will degrade and eventually be below the solar-thermal limit. If the device is too long, water vapor will evaporate from the hot base and recondense on the fins' sidewalls further downstream and decrease the device performance even further. From this analysis, the airflow velocity and ambient RH dictate the potential scale-up of 3D evaporating arrays.

**Performance of device in natural convective conditions**

The above discussion made it clear that the enhanced nominal evaporation rate of the extended fin due to environmental heat input depends on the humidity of its surrounding air. In natural convective conditions, the airflow velocities and its effects on vapor transport are strongly coupled to temperature and vapor concentration gradients due to induced buoyancy effects.[33,34] Surfaces hotter than the ambient environment will cause the air to rise upwards and form a plume. Surfaces colder than the ambient will flow downwards. If the base area is an interfacial solar absorbing material, it will become hotter than the ambient. If the base plate is reflective like white foam, the temperature will be slightly elevated, and the diffuse solar reflections from the foam base will cause the bottom region of the evaporating fin to become hotter.

The airflow velocities will depend on the geometry and the boundary layer that forms over the entire device. The natural convection velocities for single fins in these geometries are on the order of cm/s. For scaled-up devices, the airflow velocity will decrease due to the thicker boundary layers it forms (Fig. 5a). The fins near the edge of the device will have the highest heat transfer coefficients and airflow velocities due to it being adjacent to the air reservoir. The interior airflow velocities will be much lower, and it will decrease as the device's transverse width and length (x- and y-axis) increases. The sluggish airflows can lead to the hot vapor near the fins solar absorbing top and base to recondense on the colder sidewalls, as hypothesized in a previous experimental study.[31] This will strongly limit the scale-up potential of these evaporating fins in natural convective conditions.

**Closed system devices can't absorb environmental heat.** These challenges are evident in closed environment devices, such as solar still desalination devices (Fig. 5b). Enhanced performance from environmental heat absorption only occurs if the air is below 100% RH. Furthermore, it relies on continuous air supply from the environment because the air will continue to cool down and become more humid as the fins absorb environmental heat as shown in Fig. 4b and 4e. In a closed environment system, there is no net airflow and only internal circulation patterns will form. The condenser should be the coldest temperature in the system and will be at 100% RH near its surface. The hot solar evaporating surfaces are at an elevated temperature and its corresponding 100% RH as well. At steady-state, the air inside of a simple solar still device must be close to the saturated condition because it is bound by these two conditions. Unless the device's lateral dimension is so small that a non-condensing sidewall of the enclosure remains below the ambient temperature and transfers heat from outside, the only other heat inputs into the device are from the brine reservoir below and solar absorption. Thus, large-scale conventional solar stills relying on internal circulating natural convection cannot exceed the solar-thermal limit.

**FEA simulation of recondensation.** To further illustrate these challenges, we used transient finite element analysis (FEA) to simulate natural convective evaporation in a 1D straight fin geometry with sunlight normal to the z-axis ($\theta = 0°$). The highlighted yellow regions indicate the solar absorbing interfaces. This simulation will be representative of a very large system in which the interior airflow velocities become small. Fig. 5c illustrates the temperature profiles and the natural convection flow patterns that form after 10 minutes of simulation time if the base evaporates water as well. Fig. 5d illustrates the vapor molar density profile and diffusion vapor fluxes for the same case. Fig. 5c shows an internal natural convection flow pattern forming between the fins. The base plate's solar absorbing region is not as hot as the fin's tips due to heat exchange with the brine reservoir beneath it. The fin's sidewalls are at a lower temperature. However, it can be seen clearly in Fig. 5d that all the water vapor that evaporates from the base plate recondenses on the fin's sidewalls. A significant amount of vapor from the fin's top surface circulates and recondenses at the fin's sidewalls as well. This recondensation problem is still illustrated in the case of Fig. 5e and Fig 5f, which show the same profiles for the case in which the base plate does not evaporate any water. The hot tips form a high concentration vapor region at 100% RH that blocks off the sidewall's evaporation flow pathway, leading to the fin's sidewalls to only recondense water vapor. The low airflow velocities and incompatible device geometries with environmental heat absorption mechanisms lead to poor performances in scaled up arrays under natural convection.

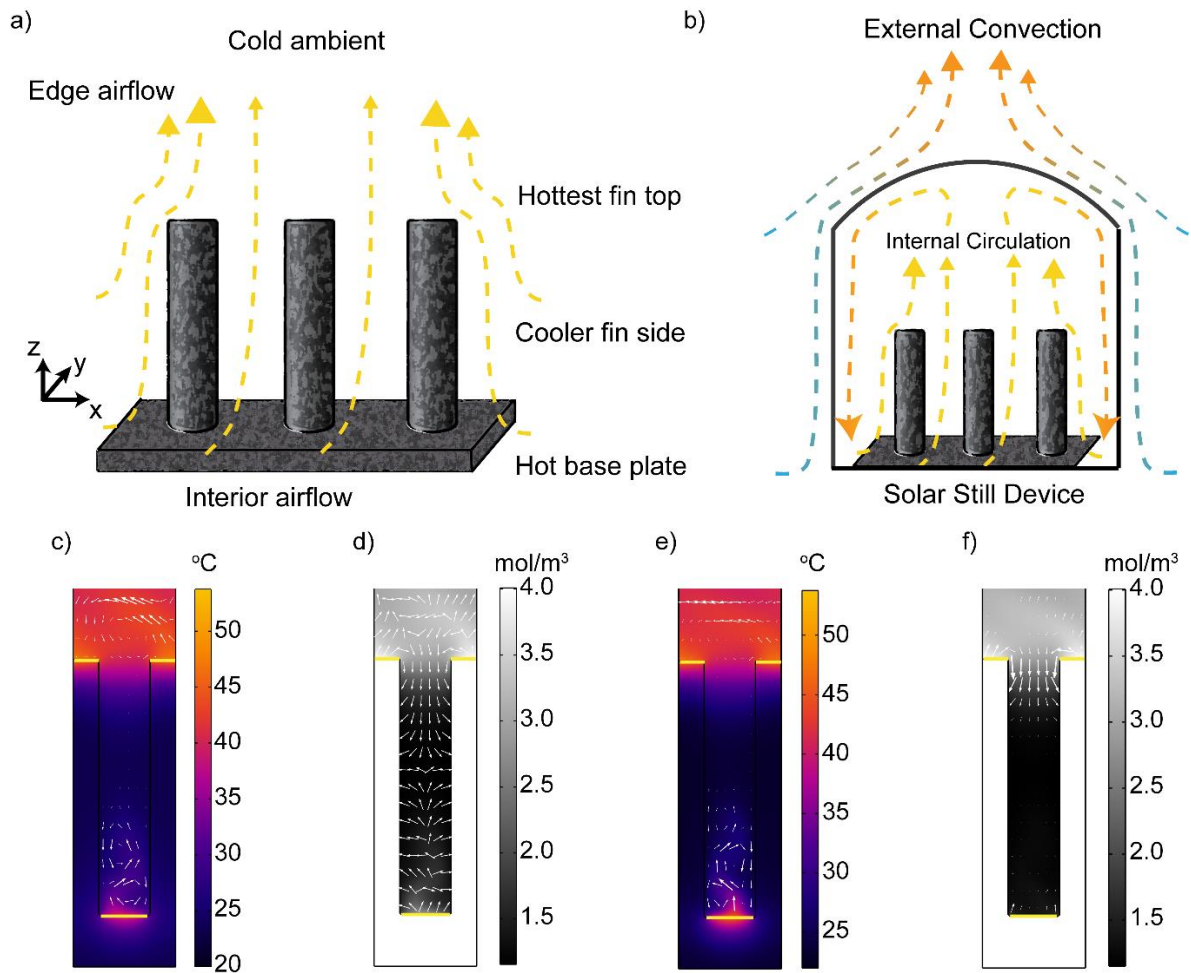

**Fig. 5** 3D fin performances in natural convective conditions. Hypothesized natural convection flow patterns in an a) open and b) closed environment such as a solar still device. Transient FEA simulated snapshots of 1D fin geometry for the case in which c)-d) the base evaporates water and e)-f) the base doesn't evaporate after 10 minutes of simulated time. c) and e) illustrate the natural convective flow patterns in white arrows and the temperature profiles. d) and f) illustrate the vapor molar density and the vapor diffusion gradients in white arrows. The outlined yellow regions illustrate the solar absorbing areas.

**Regime map of 3D device performances**

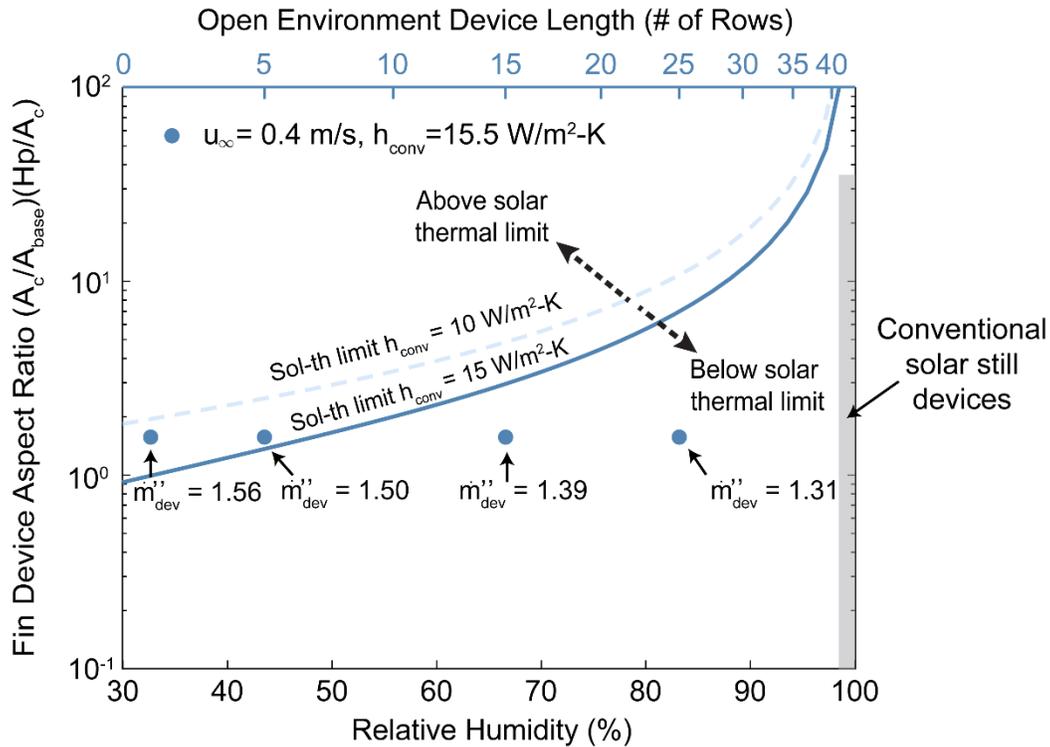

**Fig. 6** Regime map of device performances. Lines represent single fin performances equal to the solar-thermal limit from Fig. 2b. The extra area pre-factor in the y-axis account for the extra solar irradiation on the base of the device. Example forced convection device evaporation rates from Fig. 3 are reproduced to illustrate fin device's aspect ratio's ability to predict super solar-thermal evaporation rates. The length of the devices corresponding to the average RH in the device with bulk airflow $u_\infty = 0.4$ m/s is reproduced as well. The distance between each row is 5 cm. The operating region of conventional solar still devices are outlined by the gray box.

Summarizing the above discussions, we illustrate a regime map in Fig. 6 showing the required height to reach the solar-thermal limit, simulated performances of different forced convective conditions, and the region in which conventional solar still devices operate in. Unlike in Fig. 2b, an extra area ratio pre-factor $A_c/A_{base}$ is included to account for the device absorbing more solar radiation than just the fin's cross-sectional area. This allows us to map the device performances back to the single fin regime map. The predicted device evaporation rates for some scenarios in Fig. 3 are mapped into Fig. 6, showing agreement with the solar-thermal limit curves corresponding to their respective convective heat transfer coefficients. Furthermore, the predicted device evaporation rates are much lower than the single fin nominal evaporation rate performances predicted in Fig. 1e, illustrating that the nominal evaporation rate is not an accurate performance metric in array devices. The total device lengths are also reproduced in the top x-axis to illustrate the scale-up potential of these devices relative to the chosen bulk airflow velocity. Since the longitudinal spacing between rows is 5 cm, the reported devices extend up to 2 meters in length. The figure show that to achieve evaporation beyond the solar-thermal limit, the device length and the fin density can't be too high. For a given environment, there is an optimum fin spacing and size to maximize its performance.

The region in which most reported conventional solar still devices operate are shown by the gray box. Due to the device utilizing natural convection with low induced airflow velocities, the heat transfer coefficients are unlikely to exceed 10 W/m$^2$-K. The air inside of the device at steady state must be close to the saturated condition because of the working principle of the device. The regime map clearly illustrates that conventional solar still devices can't utilize environmental heat input due to the lack of net airflow through the device, the high device RH, and the low airflow velocities. The enhanced outdoor solar still performances reported previously are likely due to a combination of changing solar view factors and improper area normalizations.

**CONCLUSIONS**

3D solar interfacial-evaporation structures have been experimentally demonstrated to be able to evaporate water exceeding the solar-thermal limit due to environmental heat absorption. Although many lab-scale devices have been built and tested, there exist no criteria distinguishing 2D from 3D structures and very few attempts to test and analyze their scale-up potential for large scale deployment.

We have systematically modeled and explained the physics underlying the performance of 3D evaporating structures. First, the commonly reported nominal evaporation flux and the device evaporation fluxes are distinguished from each other due to the different area normalizations. A single fin model was constructed and was successfully able to reproduce the flat temperature profiles observed in experiments as well as the enhanced nominal evaporation rates. The model predicts that the nominal evaporation rate scales linearly with the fin height and that the environmental heat input critically relies on the RH to be below the saturation point. Using the model, we have identified the non-dimensional critical heights in which a fin will first begin to absorb environmental heat and it absorbs enough environmental heat to nominally evaporate at the solar-thermal limit.

The single fin model was extended to a coarse-grained fin array to study the scale-up performance of these devices by coupling with the vapor and heat transport on the air side as well. Using this model, we have highlighted the potential and limitations in increased device performances by adding fin structures in forced convection situations. The model was able to illustrate the difference in performance between solar and dark evaporation, the environmental heat absorption, and the non-monotonic local air temperature profiles. The model pinpointed that the common procedure of using a foam insulating base plate is ill-advised because the base area can contribute a significant amount of evaporation flux due to enhanced kinetics from solar absorption. The model illustrates that the greatest gain in performance can occur if the external airflow velocity is high so that it can enhance the convective mass transfer coefficients and quickly replenish the humid air inside of the device with dry air. The device width can only be extended up to when the air becomes fully saturated with vapor. Afterwards, device performance will degrade significantly because hot vapor from the solar absorbing regions can recondense on the cooler fins' sidewalls.

We have identified that the enhanced benefits from 3D evaporating fins are unlikely to translate in large devices for natural convection open environment conditions due to the low airflow velocities. We have also illustrated that environmental heat absorption cannot be the mechanism for enhanced performance in natural convection closed environment conditions, such as in solar still devices, because the air internally is close to 100% RH. Instead, these small-scale devices have misleading improved performances due to the changing solar view factors on the fin and improper area normalization. Through FEA simulations, we have further clarified and illustrated the recondensation effect in natural convective conditions due to the hot vapor region that forms from the evaporating tip, blocking off the vapor transport pathway from the sideways.

We have compiled these findings into a regime map to illustrate the performances of single fins, example simulated studies of scaled-up forced convection devices in open environment, and the regime in which conventional solar stills operate in. An extra area ratio pre-factor allows us to approximately map back the overall device performance to the predicted performances of a single fin.

We conclude this paper by emphasizing the fundamental difference between 2D and 3D solar interfacial-evaporators with evaporation rates surpassing the solar-thermal limit. As shown in this paper, a 3D evaporator can absorb heat from the environmental and achieving evaporation rates exceeding the solar-thermal limit.[13,14] In this case, the evaporation is still thermally driven. No matter if the evaporators are 2D or 3D, if nowhere in the system is below the ambient temperature, it is impossible to exceed solar-thermal limit based on purely thermal processes because all the heat used for evaporation must come from the solar energy. For 2D evaporators, it is difficult to achieve below ambient temperature unless there is a very high air flowrate.[18] Thus, most 2D evaporators tested under natural convection condition with evaporation rates exceeding the solar-thermal limit implies a non-thermal evaporation process, such as via the photomolecular effect that we have discussed in several publications.[8,20–22] We hope this work, together with our investigations on 2D evaporators, will provide stimulus for future understanding of the mechanisms of solar interfacial evaporators exceeding the solar-thermal limit, which will enable better applications.

**METHODS**

**Single fin model**

We make the following key assumptions in constructing an ideal performance model for a single fin: (1) we will ignore the detailed capillary flow of liquid brine in the 3D evaporator, (2) the external surface of the evaporating structure are always fully wetted, (3) the device will have a mechanism to reject salt accumulation during evaporation to prevent salt crystallization, and (4) the properties of the brine solution can be approximated using the properties of pure water. These assumptions are made to allow a systems level modeling of the device while having results that are generalizable to different materials system. Assumption 1 holds because the net mass flux of brine is small due to the low evaporation rates of 1 to 10 kg/$m^2$-h. Assumption 2 requires that the 3D solar evaporator have enough capillary pumping to continuously supply water without drying out and would depend ultimately on the material's pore structure, surface energy, and height of the fin. Assumption 3 is the least rigorous assumption due to the low diffusivities of ions and water relative to the evaporation rate and would place limits on the device geometry, leading to these results becoming a best-case scenario. Assumption 4 neglects the small change of the colligative properties such as vapor pressure depression due to the presence of salts. Thus, assumptions 2-4 govern the "ideal" performance of a 3D evaporator and the relaxation of each assumption will lead to performance degradation.

With such approximations, the differential energy balance equation can be written as the sum of conductive heat transfer along the fin and the convective, evaporative, and radiative heat exchange with the ambient reservoir.

$$0 = A_c k_m \frac{d^2 T}{dz^2} - p\left(h_{conv}(T - T_\infty) + h_{fg} g_{evap} C_g \left(c_{v,s}(T) - RH c_{v,s}(T_\infty)\right) + \sigma \epsilon (T^4 - T_\infty^4)\right) \quad (10)$$

where $T$ is the local fin temperature, $T_\infty$ is the ambient temperature, RH is the ambient relative humidity, $z$ is the axis along the height of the extended surface, $A_c$ is the cross-sectional area of the extended fin, $k_m$ is the wetted material's thermal conductivity, $p$ is the perimeter, $h_{conv}$ is the convective heat transfer coefficient, $h_{fg}$ is the latent heat of evaporation, $g_{evap}$ is the convective mass transfer coefficient, $c_{v,s}$ is the saturated molar vapor fraction at a given temperature, $\sigma$ is the Stefan-Boltzmann constant, and $\epsilon$ is the blackbody emissivity of the wetted 3D evaporator. The boundary conditions of this equation are two Robin style boundary conditions governing the heat fluxes. At the top interface, it will exchange heat and vapor with the ambient environment and absorb sunlight.

$$q(z = H) = A_c \left(q_{sun}'' - h_{conv,top}(T - T_\infty) - h_{fg} g_{evap,top} C_g \left(c_{v,s}(T) - RH c_{v,s}(T_\infty)\right) - \sigma \epsilon (T^4 - T_\infty^4)\right) \quad (11)$$

The heat and mass transfer correlations are described using the heat and mass transfer analogies for crossflow on a cylinder (for side wall) and laminar flow on a flat plate (for top surface). For crossflow, we will use the Churchhill correlation.[33,35] The corresponding Nusselt numbers (Nu), and their relationships to the convective transfer coefficients are

$$Nu_{conv,side} = 0.3 + \frac{0.62 Re_D^{1/2} Pr^{1/3}}{(1 + (0.4/Pr)^{2/3})^{1/4}} \quad (12)$$

$$Nu_{conv,top} = 0.664 Re_D^{1/2} Pr^{1/3} \quad (13)$$

$$h_{conv} = \frac{Nu\, k_{air}}{L_D} \quad (14)$$

where $Pr$ is the Prandtl number of air, $Re$ is the Reynolds number, and $k_{air}$ is the thermal conductivity of air. We used the same correlations to describe the mass transfer of vapor in air. The mass transfer coefficient can be found analogously using Eq. (12-14) by replacing $h_{conv}$ with $g_{evap}$, $Nu$ with the Sherwood number $Sh$, $Pr$ with the Schmidt number $Sc$, and $k_{air}$ with the diffusion coefficient of water vapor in excess of air $D_v$. The bottom heat transfer coefficient, $h_{bot}$, governs the convective heat transfer between the brine pond reservoir and the evaporating fin's bottom interface. The true value should depend on the temperature differences and the convective flow patterns inside of the reservoir. For simplicity, we will set $h_{bot}$ to 100 W/m²-K and the brine pond $T_b$ to the same temperature as the ambient $T_\infty$ at 23 ºC. To account for the foam inserts used in the base in laboratory tests or the base plate thickness in outdoor devices, we will assume that there is an additional $t_{base}$ of 2 cm of wetted material not exposed to the ambient environment. The bottom Robin boundary condition is thus

$$q(z=0) = \frac{A_c(T_b - T)}{1/h_{bot} + t_{base}/k_m} \tag{15}$$

A system of non-linear equations is formed that relates the heat fluxes at each local cross-section of the fin at a given iteration n.

$$\bar{f}_n = \bar{\bar{A}}\bar{T}_n + \bar{b}_n + \bar{c} \tag{16}$$

where $\bar{f}_n$ is the function vector that becomes zero at the equilibrium temperature profile, the $\bar{\bar{A}}$ matrix has constant terms that couple linearly with the temperature vector $\bar{T}_n$, the $\bar{b}_n$ vector has terms that are non-linearly dependent on the temperature, and the $\bar{c}$ vector holds constant terms. The Jacobian can then be calculated by taking the multivariable derivative of each term in Eq. (16).

$$\bar{\bar{J}}_n = \bar{\bar{A}} + \frac{\partial \bar{b}_n}{\partial \bar{T}} \tag{17}$$

The newly guessed temperature profile can then be calculated using the Jacobian.

$$\bar{T}_n^* = \bar{T}_n - \bar{\bar{J}}_n^{-1}\bar{f}_n \tag{18}$$

The new temperature profile is then iterated by mixing part of the newly guessed temperature $\bar{T}_n^*$ and the old temperature $\bar{T}_n$ using a mixing term $\lambda$.

$$\bar{T}_{n+1} = \lambda \bar{T}_n^* + (1-\lambda)\bar{T}_n \tag{19}$$

**Coarse-grained scaled-up fin model**

A coarse-grained model is developed using the single fin results to build the scaled-up model for fin arrays. Further simplifications are imposed to make the model computationally tractable. The first assumption is that due to the complex view factors and temperature profiles internal to the pin fin array, we will neglect radiation heat exchange. This effect will cause the temperature difference between the hot and cold evaporating surfaces to be larger, however we expect this effect to become small as the external airspeed and convective heat transfer coefficients increase. The second assumption is that there is a clear separation between air flow that is "internal" (flowing through the pin fin array) and a free stream "external" air flow on the tips of the pin fin. This separation will clearly define the boundary condition for the fins' tips and the internal flow that will progressively become more humid. This simplification will likely over-predict the evaporation rate from the top cross-sectional area because this region will couple with the internal flow and become more humid locally as well. The third assumption is that due to the various pins dispersed throughout the array, it will induce local mixing and allow us to describe the heat and mass transfer using the bulk-averaged properties of the airflow at each control volume. The fourth assumption is that the air always flows in the same direction and that the array is large enough in the transverse direction (x-axis) so that there are minimal edge effects. The final assumption is that the temperature of the base plate in each control volume can be estimated independently of its neighbors using a 1D heat and mass transfer resistance network of a flat plate correlation in which there are no pin fin array. This assumption becomes more valid as the spacing between each fin becomes large relative to the fin's characteristic size and the base plate thickness.

The model works by doing a control volume analysis of air flowing downstream along the device and coarse graining the total heat and mass transfer exchange between it and the local row. At the inlet of the ith-row pin fin array, we denote the air temperature $T_{air,i}$, vapor mole fraction $c_{v,i}$, and velocity. The total molar flowrate of "internal" air is $\dot{m}_{air} = C_g u_\infty S_t H$. Žukauskas correlation for tube bundles are used to describe heat transfer between the fins and the air.[33,36,37]

$$Nu_{conv} = Pr^{0.36} fn(Re_{L_c, u_{max}}) \qquad (20)$$

$$fn(Re_{D,u_{max}}) = \begin{cases} 0.71 Re_{L_c, u_{max}}^{0.5}, & Re_{D,u_{max}} < 1180 \\ 0.35 Re_{L_c, u_{max}}^{0.6}, & Re_{D,u_{max}} \geq 1180 \end{cases} \qquad (21)$$

where $u_{max}$ is[33]

$$u_{max} = u_\infty \frac{S_t}{S_t - L_c} \qquad (22)$$

due to the venturi effect. The corresponding Sherwood numbers and mass transfer coefficients are found using the heat and mass transfer analogy again. To describe the heat ($h_{base}$) and mass transfer ($g_{evap,base}$) coefficients between the base plate and the air, the flat plate correlations in Eq. (13) is used with a characteristic size of $S_t$. The base temperature and evaporation rates are then calculated by using a 1D heat and mass transfer resistance network at the evaporating surface.

$$0 = q''_{sun} - h_{fg} C_g g_{evap,base}(c_{v,s}(T_{base,i}) - c_{v,i}) - h_{base}(T_{base,i} - T_{air,i}) - \frac{(T_{base,i} - T_b)}{1/h_{bot} + t_{base}/k_m} \qquad (23)$$

As the airflows over each row, the air's temperature and vapor content will change based on the evaporation and heat exchange.

$$c_{v,i+1} = c_{v,i} + \frac{(S_t S_l - A_c) g_{evap,base} C_g (c_{v,s}(T_{base}) - c_{v,i}) + P g_{evap,side} C_g \int_0^H (c_{v,s}(T) - c_{v,i}) dz}{\dot{m}_{air}} \qquad (24)$$

$$T_{air,i+1} = T_{air,i} + \frac{(S_t S_l - A_c) h_{base}(T_{base} - T_{air,i}) + P h_{conv,side} \int_0^H (T - T_{air,i}) dz}{\dot{m}_{air} c_p} \qquad (25)$$

Heat conduction and vapor diffusion along the air profile is neglected due to the high Peclet numbers involved in the study. The profiles are then solved by forward marching in space along the array length using the updated humidity and temperatures of the air.

**FEA model**

We have modeled natural evaporation from the 1D fin array using transient simulations in a 2D geometry. Due to the large-scale nature of these devices, we have modeled the behavior of the interior fins, which should dominate the overall device behavior. Periodic boundary conditions are imposed in the perpendicular directions for heat transport, vapor transport, and momentum equations. The top of the domain is set to be 10 times taller than the fin and set as an open boundary condition. Air is fully simulated around the fin structure. The general governing equations for mass, momentum, energy, and vapor transport in the air are

$$\frac{\partial \rho}{\partial t} + \nabla \cdot (\rho \vec{u}) = 0 \tag{26}$$

$$\rho \frac{\partial \vec{u}}{\partial t} + \rho (\vec{u} \cdot \nabla) \vec{u} = -\nabla p + \nabla \cdot \left( \mu (\nabla \vec{u} + (\nabla \vec{u})^T) - \frac{2}{3} \mu (\nabla \cdot \vec{u}) I \right) + \rho \vec{g} \tag{27}$$

$$\rho C_p \left( \frac{\partial T}{\partial t} + \vec{u} \cdot \nabla T \right) = \nabla \cdot (k \nabla T) \tag{28}$$

$$\frac{\partial c_v}{\partial t} + \nabla \cdot (c_v \vec{u}) = \nabla \cdot (D_v \nabla c_v) \tag{29}$$

Using the weakly compressible mode, air's density is coupled with its temperature to induce natural convective flows. Only heat conduction is modeled inside of the array structure due to the low flowrates of liquid involved. The material's thermophysical properties are set to a thermal conductivity of 0.3 W/m-K, a specific heat capacity of 2000 J/kg-K, and a density of 1200 kg/m$^3$. The fin was set to 10 cm tall, the base plate set to 2 cm in thickness, the width of the fin set to 2 cm, and the gap between the fins is set to 2 cm. The bottom of the baseplate is set to a Robin heat boundary condition through a heat transfer coefficient of 100 W/m$^2$-K interacting with a reservoir temperature of 23 °C. The air surrounding the fin is initially set to 23 °C and 30% RH. The top of the fins and the top of the base plate absorbs 1000 W/m$^2$ to simulate solar absorption with a zenith angle of 0°. The simulation is then forward marched in time to produce the simulation snapshots.

**Author contributions**

J. H. Z. and G. C. developed the concept. J. H. Z. prepared the models. J. H. Z., R. M., A. O., and G. C. wrote the paper. G. C. directed the overall research.

**Conflicts of interest**

There are no conflicts to declare.

**Data availability**

The modeling methods and formula have been fully included in the Methods section.

**Acknowledgements**

We gratefully acknowledge funding support from the Abdul Latif Jameel Water and Food Systems Lab (J-WAFS), UM6P & MIT Research Program (UMRP), and MIT Bose Award. J. H. Z. acknowledges support from the J-WAFS Graduate Student Fellowship, the MathWorks Mechanical Engineering Fellowship, and the Harriet and Chee C. Tung Global Collaborative Fellowship. The authors would like to acknowledge Dr Wenhui Tang for her feedback on the figures.